\documentstyle[12pt]{article}
\newcommand{\be}{\begin{equation}}
\newcommand{\ee}{\end{equation}}
\newcommand{\ba}{\begin{eqnarray}}
\newcommand{\ea}{\end{eqnarray}}
\newcommand{\ed}{\end{document}}
\newcommand{\lab}[1]{\label{#1}}
\newcommand{\re}[1]{(\ref{#1})}
\newcommand{\ci}[1]{\cite{#1}}
\newcommand{\bfr}{\begin{flushright}}
\newcommand{\efr}{\end{flushright}}
\newcommand{\bfl}{\begin{flushleft}}
\newcommand{\efl}{\end{flushleft}}
\renewcommand{\baselinestretch}{1.4}
\date{}

\title{CHAOTIZATION OF THE SUPERCRITICAL  ATOM}
\author{ D.U.MATRASULOV \\
 Heat Physics Department of the Uzbek Academy of Sciences,\\
28 Katartal St.,700135 Tashkent, Uzbekistan }
\begin{document}\large
\maketitle

\begin{abstract}
Chaotization of supercritical ($Z>137$) hydrogenlike
atom in the monochromatic field is investigated. A theoretical analysis
of chaotic dynamics of the relativistic electron based on Chirikov
criterion is given.
Critical value of the external field at which chaotization will occur
is evaluated analytically. The diffusion
coefficient is also calculated.
\end{abstract}

Study and synthesis of superheavy elements is becoming one of actual problems
of the modern physics \ci{per98,hol99}. Fast growing interest to the physics
and chemistry of actinides and transactinides stimulates extensive study
of superheavy elements. One of the main differences which leads to the additional
difficulties in the study of superheavy atoms is the fact that the motion of
the atomic electrons is described by the relativistic equations of motion
due to the large values of the charge of the atomic nucleus. In this Brief Report
we will study classical chaotic dynamics of the relativistic hydrogenlike atom
with the charge of the nucleus $Z>137$, interacting with monochromatic field.
Such an atom is called the overcritical atom \ci{mur78a}-\ci{mig77}.
Quantum mechanical properties of such atom was investigated by a number of
authors \ci{pom45,piep69,mig77}.
Quasiclassical dynamics of the supercritical atom was investigated by
V.S.Popov and co-workers \ci{mur78a}-\ci{mur78b}.
The experimental way of creating the overcritical states are the collision
experiments of slow heavy ions with resulting charge $Z_1+Z_2>137$ \ci{Grib,pop73,raf78}.
To treat chaotic dynamics of the relativistic electron in the
supercritical kepler field we need to write the unperturbed Hamiltonian
in terms of action-angle variables.
As is well known \ci{mur78a}, for the relativistic electron moving in the
field of charge $Z>137$ point charge approximation cannot be applied for
describing its motion i.e. there is need in regularizing of the problem.
Such a regularizing can be performed by taking into account a finite sizes
of the nucleus i.e. by cut off  Coulomb potential at small distances:\\
$$
V(r)=\left\{\begin{array}{ll}-\frac{Z\alpha}{r}\;, & for\;\:r>R \\ \\
-\frac{Z\alpha}{b}f(\frac{r}{R})\,, & for\;\;0<r<R\end{array} \right.\,, $$

where  $f(\frac{r}{R})$ is the cut-off function
$R$ is the radius of the nucleus, $\alpha = 1/137$ (the system of units $m_e =\hbar=c=1$ is used here and below).
Further we will take $f(r/R)=1$
(surface distribution of the charge). Then the relativistic momentum defined by
$$
p = \sqrt{(\varepsilon -V)^2 -\frac{M^2}{r^2}- 1},
$$
where $\varepsilon$ is the energy of the electron and  $M$ its angular
momentum, can be rewritten as the following:
$$
p=\left\{\begin{array}{ll} \sqrt{(\varepsilon + \frac{Z\alpha}{r})^2 -\frac{M^2}{r^2}- 1} \; & for\;\:r>R \\ \\
\sqrt{(\varepsilon + \frac{Z\alpha}{R})^2 -\frac{M^2}{r^2}- 1}\; &for\:\;0<r<R\end{array} \right.\,, $$

One of the turning points of the electron (which are defined as a zeros of the
momentum) lies on inside of the nucleus and given by
$$
r_1 = M[(\varepsilon +\frac{Z}{R})^2 - 1]^{-1}
$$

The turning point lying on outside of the nucleus is given by the expression
$$
r_2 = \frac{\mid\varepsilon Z\mid-\sqrt{\varepsilon^2Z^2-(\varepsilon^2-1)
(Z^2-M^2)}}{\varepsilon^2-1}
$$

Thus one can write for the action (for  $Z>M$)
\be
\pi n = I_1+I_2,
\lab{non}
\ee

where
$$
I_1 = \int_{r_1}^{R}\sqrt{(\varepsilon +\frac{Z}{R})^2-\frac{M^2}{r^2}-1}\; dr
$$

$$
I_2 = \int_{r_1}^{R}\sqrt{(\varepsilon +\frac{Z}{r})^2-\frac{M^2}{r^2}-1}\; dr
$$

From \re{non} one can find the Hamiltonian of the relativistic electron in the
field of overcritical nucleus ($Z>137$) in terms of action-angle variables:
\be
H_0 = \varepsilon \approx -\frac{g}{Z}c(R,g)exp\{-\frac{\pi n}
{g}\},
\lab{nonp}
\ee
where $c(r,g)= exp(g-R)$.

In the derivation of \re{nonp} we have accounted that $Z\sim M$ and $\varepsilon \sim 0$.
The Kepler frequency can be defined as
\be
\omega_0 = \frac{dH_0}{dn}= \frac{\pi}{Z}c(R,g)exp\{-\frac{\pi n}{g}\}
\ee

Trajectory equation for $Z>M$ (for $r>R$) has the form \ci{land}
\be
\frac{Z^2-M^2}{r} = \sqrt{M^2\varepsilon^2 +(Z^2-M^2)} ch(\phi\sqrt{\frac{Z^2}{M^2}-1})
+\varepsilon Z
\ee

For $Z\sim M$ ($\varepsilon\sim 0$) we have
$$
\frac{g^2}{r} \approx \sqrt{M^2\varepsilon^2 +g^2}(1- \phi^2)\sqrt{\frac{Z^2}{M^2}-1}
+\varepsilon Z
$$
or
$$
\frac{g}{r} \approx g(M- \phi g) +\varepsilon Z
$$

The trajectory equation for  $r<R$ is
\be
\frac{1}{r} = a_0 cos\phi,
\ee
where
$$
a_0 = M^{-1}[(\varepsilon +\frac{Z}{R})^2 - 1]
$$
To investigate the chaotic dynamics of this atom we will consider the
angular momentum as fixed and ($Z\approx M$).

Consider now the interaction of the supercritical atom with a linearly polarized
monochromatic field
\be
V = \epsilon cos\omega t sin\theta[xsin\psi +ycos\psi],
\ee

where $\theta$ and $\psi$ are the Euler angles.

The full Hamiltonian of the system can be written as
\begin{eqnarray}
\displaystyle
H = -\frac{\sqrt{Z^2-M^2}}{Z}exp\{-\frac{\pi n}{\sqrt{Z^2-M^2}}\} +
 \nonumber\\ \nonumber\\
\epsilon cos\omega t sin\theta\sum(x_ksin\psi cosk\lambda +
y_k cos\psi sink\lambda),
\lab{lin}
\end{eqnarray}

where $x_k$ and $y_k$ are the Fourier components of the electron
dipole moment:
\begin{eqnarray}
\displaystyle
x_k = \frac{i}{k\omega T}\int_{0}^{T}e^{ik\omega t} dx =\frac{i}{k\omega T}\int_{0}^{T}
e^{ik\omega(2\mid\varepsilon Z\mid\xi-sin\xi)} \nonumber\\
sin\xi\{cos\frac{2g^{-1}}{a-cos\xi}-\frac{1}{a-cos\xi}sin\frac{2g^{-1}}{a-cos\xi}\}d\xi
\lab{xint}
\end{eqnarray}
and
\begin{eqnarray}
\displaystyle
y_k = -\frac{iMb^{-5/2}}{k\omega T}\int_{0}^{T}
\frac{e^{ik\omega t} sin2\xi d\xi}{\sqrt{M^2cos^2\xi - b^2}} +\frac{i}{k\omega T}\int_{0}^{T}
e^{ik\omega(2\mid\varepsilon Z\mid\xi-sin\xi)} \nonumber\\
sin\xi\{sin\frac{2g^{-1}}{a-cos\xi}-\frac{1}{a-cos\xi}cos\frac{2g^{-1}}{a-cos \xi}\}d\xi
\lab{yint}
\end{eqnarray}

here $a = \sqrt{Z^2-M^2}exp\{-\pi n/\sqrt{Z^2-M^2}\},\;\;\;$
$T = 2\pi/\omega_0$ $$b = (\varepsilon +\frac{Z}{R})^2 -1,$$
$$
T_1 = \frac{(R-r_0)}{2M},\;\;\;\; T_2 = T - T_1
$$

Calculating the integrals \re{xint} and \re{yint} using the stationary phase
method we have
\be
x_k = 0, \;\;\;\;\; y_k = \frac{R^2 exp\{\frac{\pi n}{\sqrt{Z^2-M^2}}\}}{\pi k^2}
\lab{comp}
\ee

For the further treatment of the chaotic dynamics of the system one should
find as it was done in \ci{del83,mat1}, resonance width:
$$
\Delta\nu_k = (8\omega_0'r_k\epsilon)^{\frac{1}{2}},
$$
where $r_k = \sqrt{x_k^2+y_k^2}$.

Application of the Chirikov criterion to the Hamiltonian \re{lin}
gives us the critical value of the external field at which electron moving
in the supercritical Kepler field enters chaotic regime of motion:
\be
\epsilon_{cr} = \frac{gc(R,g)exp\{-\pi n/g\}}
{20Zk(k+1^2)(\sqrt{r_k}+\sqrt{r_{k+1}})^2}
\ee

Taking into account \re{comp} for the critical field we have
\be
\epsilon_{cr} = \pi k \frac{gc(R,g)exp\{-2\pi n/g\}}
{20Z(2k^2+2k+1)}
\lab{overcr}
\ee

One can also calculate the diffusion coefficient:
\be
D = \frac{\pi}{2}\frac{\epsilon^2R^2}{c(R,g)Z^3}exp\{2\pi n/g\}
\ee

Thus we have obtained the critical value of the external monochromatic
field strength at which chaotization of motion of the electron moving in the
supercritical Kepler field will occur. As is seen from \re{overcr}
the this critical value is rather small i.e. in the supercritical case
($Z>137$) electron is more chaotic than the undercritical ($Z<137$) case.
This can be explained by the fact that the level density of the $Z>137$
atom is considerably more (see \re{nonp}) than the one for the undercritical
atom (see \ci{mat1}). The above results may be useful for slow collision
experiments of heavy (with the resulting charge $Z_1+Z_2>137$) ions in the presence
of laser field.

\ed